\documentclass[12pt]{article}

\usepackage[latin2]{inputenc}
\usepackage{amsmath}
\usepackage{amssymb}
\usepackage{amsfonts}
\usepackage{stmaryrd}
\usepackage{theorem}
\usepackage{graphicx}
\usepackage{epstopdf}
\usepackage[parfill]{parskip} 
\usepackage{hyperref}
\hypersetup{pdftex,colorlinks=true,allcolors=blue}
\usepackage{hypcap}
\usepackage{algorithm}
\usepackage{algpseudocode} 
\usepackage{authblk}
\usepackage{tabto}
\usepackage{wrapfig} 
\usepackage{siunitx} 

\newcommand{\keywords}[1]{\par\noindent{\small{\em Keywords\/}:\ \ #1}}
\newcommand{\mscclass}[1]{\par\noindent{\small{\em MSC class\/}:\ \ #1}}

\newcommand*{\me}{\mathrm{e}}

\newcommand*{\vr}{\ensuremath{\varrho}}

\newcommand*{\vt}{\ensuremath{\vartheta}}

\newcommand*{\vs}{\ensuremath{\sigma}}

\newcommand*{\vg}{\ensuremath{\gamma}}

\newcommand*{\R}{\ensuremath{\mathbb{R}}}

\newcommand*{\mto}{\ensuremath{\rightarrow}}

\newcommand*{\md}{\mathrm{d}} 
\newcommand{\libeq}{\mathrel{\mathop:}=} 

\newcommand*{\kon}{\ensuremath{k_{\mathrm{on}}}}
\newcommand*{\koff}{\ensuremath{k_{\mathrm{off}}}}
\newcommand*{\Kd}{\ensuremath{K_{\mathrm{d}}}}
\newcommand*{\meq}{\ensuremath{\mathrm{eq}}}
\newcommand*{\mdis}{\ensuremath{\mathrm{dis}}}

\newcommand*{\dcdt}{\ensuremath{\partial_t c}}

\newcommand*{\ddt}{\ensuremath{\frac{\mathrm{d}}{\mathrm{d}t}}}
\newcommand*{\delt}{\ensuremath{\Delta t}}
\newcommand*{\delx}{\ensuremath{\Delta x}}

\newcommand*{\dcdx}{\ensuremath{\partial_x c}}
\newcommand*{\ddcdx}{\ensuremath{\partial_x^2 c}}

\newcommand*{\Leq}{\ensuremath{\bar{L}}}
\newcommand*{\Teq}{\ensuremath{\bar{T}}}

\newcommand*{\Ceq}{\ensuremath{\bar{C}}}
\newcommand*{\Lini}{\ensuremath{L_{\mathrm{ini}}}}
\newcommand*{\Tini}{\ensuremath{T_{\mathrm{ini}}}}

\newcommand*{\tmax}{\ensuremath{t_{\mathrm{max}}}}
\newcommand*{\xdet}{\ensuremath{x_{\mathrm{det}}}}
\newcommand*{\mA}{\mathrm{A}}
\newcommand*{\mS}{\mathrm{S}}

\newcommand*{\mIC}{\mathrm{IC}}

\newcommand*{\mmet}{\ \si{\metre}}
\newcommand*{\msec}{\ \si{\second}}
\newcommand*{\mmcmol}{\ \si{\cubic\metre/\mole}}
\newcommand*{\mmsVsec}{\ \si{\square\metre/\volt\second}}
\newcommand*{\mmssec}{\ \si{\square\metre/\second}}
\newcommand*{\mmolmc}{\ \si{\mole/\cubic\metre}}

\hypersetup{bookmarksnumbered}

\begin{document}

\title{Estimating Kinetic Rate Constants\\ and Plug Concentration Profiles from\\ Simulated KCE Electropherogram Signals}
\author{J\'{o}zsef Vass,\ \ Sergey N. Krylov\footnote{Corresponding author.}}
\affil{jvass@yorku.ca,\ skrylov@yorku.ca\\ \ \\ Centre for Research on Biomolecular Interactions\\ Department of Chemistry, York University\\ Toronto, ON, M3J 1P3, Canada}
\date{April 27, 2017}
\maketitle

\begin{abstract}
\noindent Kinetic rate constants fundamentally characterize the dynamics of the chemical interaction of macromolecules, and thus their study sets a major direction in experimental biochemistry. The estimation of such constants is often challenging, partly due to the noisiness of data, and partly due to the theoretical framework. We present novel and qualitatively reasonable methods for the estimation of the rate constants of complex formation
and dissociation in Kinetic Capillary Electrophoresis (KCE). This also serves our broader effort to resolve the inverse problem of KCE, where these estimates pose as initial starting points in the non-linear optimization space, along with the asymmetric Gaussian parameters describing the injected plug concentration profiles, which we also hereby estimate. We also compare our rate constant estimation method to an earlier one, also devised by our research team.\\
\mscclass{92C45 (primary); 62H12, 92C40 (secondary).}
\end{abstract}

\newpage
\tableofcontents

\newpage
\section{Introduction} \label{s01}

The physical model of Kinetic Capillary Electrophoresis (KCE) was introduced by our lab in articles \cite{berezovski2002nonequilibrium, petrov2005kinetic} and surveyed in \cite{krylov2007kinetic}. The main purpose of the model is to enable the experimenter the reliable determination of the kinetic rate constants of complex formation and dissociation, denoted $\kon$ and $\koff$. Under specific fixed initial conditions, these two constants induce parametrized signals, arising from the solution of a constant--convection--diffusion--reaction equation (CCDR) \cite{vasskrylov2016ccdr}. The measured experimental signal poses a significant reduction in information, as it is merely the superposition of two concentration components, while the CCDR equation has actually three such components -- i.e. one component is lost entirely in the signal acquisition, while the other two are summed, as explained in the next section.

Despite the complexity of this estimation task, hinted above, our team has introduced an integration-based method, which has gone through several stages of improvement, and we review the latest version in Section \ref{s0202}, along with the series of papers that led up to it. The crux of the method is determining the three disjoint subintervals of the experimental time interval, in which the signal acquisition occurs at the detector location. Two of these three intervals correspond to a left- and right-peak of the superposed concentration functions, mentioned earlier, while the third is the dissociation bridge. The area under the signal over these three intervals imply the area-based estimates for $\kon$ and $\koff$.

Our new estimation method introduced in Section \ref{s03} also requires the above three intervals, but the requirement of accurate integrability is replaced with a more robust linear least squares regression, executed either directly on the signal data sample, or data derived via deconvolution. The former is the case for our estimation of $\koff$, while the latter for that of the asymmetric Gaussian plug parameters, which imply our estimate of $\kon$. The exact solutions of KCE for a simplified case, derived in \cite{vasskrylov2016ccdr}, are utilized towards this estimate. The accuracy of our rate constant estimates is compared to the area-based method in Section \ref{s0402}, while the plug parameter estimates are analyzed in Section \ref{s0403}.

To give a brief survey of the relevant literature, firstly we observe that various experimental approaches have been taken to approximating the rate
constants of complex formation and dissociation, $\kon$ and $\koff$. Methods by others are \cite{wilson2002analyzing, hornblower2007single, al2005fluorescence, li2007kinetics, abdiche2008determining, rich2007higher}, while the KCE-based methods by our lab are surveyed in Section \ref{s0202}. Each method has its advantages and limitations, while their applications also vary. For instance, Hornblower et al. \cite{hornblower2007single} present a nanopore amperometric approach, which requires the microscopic observation of a few molecules, similarly to another approach via fluorescence correlation spectroscopy \cite{al2005fluorescence, li2007kinetics}. Other approaches, such as one via surface-immobilized binding sensors \cite{abdiche2008determining, rich2007higher}, or our own via KCE, are macroscopic since they require measuring changes in concentrations. KCE methods \cite{krylov2007kinetic} have their advantage in measuring rate and equilibrium constants with only a relatively small amount of the required substances, which are often expensive, on the condition that one of the reactants can be labeled fluorescently, without affecting the interaction.

\newpage
\section{Preliminaries} \label{s02}

\subsection{The Physical Model} \label{s0201}

Our previous article \cite{vasskrylov2016ccdr} describes how the KCE equations \cite{krylov2007kinetic, berezovski2002nonequilibrium, petrov2005kinetic} originate in the Nernst--Planck Equations, which model the combined effects of convection, diffusion, and chemical reaction between ions in some electric field.

The concentration vector of three reactants $c = (L, T, C):\R_+^2\mto\R_+^3$ denoting the ligand, target, and complex, is defined over spacetime points $(t,x)\in [0, \tmax]\times [0, \xdet]$, where the measurement occurs at the detector $\xdet$. The concentrations satisfy the equation
\[ \dcdt + v\cdot\dcdx = D\cdot\ddcdx + R(c) \]
where $v = (v_L, v_T, v_C)\in\R_+^3$ and $D = (D_L, D_T, D_C)\in\R_+^3$ are the constant velocity and diffusion vectors, and $\cdot$ denotes the Hadamard product. The reaction term takes the form
\[ R(c) = (-\kon LT + \koff C,\ -\kon LT + \koff C,\ \kon LT -\koff C): \R_+^2\mto\R^3 \]
where $k = (\kon,\ \koff)\in\R_+^2$ are the rate constants of complex formation
and dissociation respectively. The equilibrium dissociation constant is defined as $\Kd\libeq\koff/\kon$.

This system of partial differential equations must be accompanied by appropriate initial and boundary conditions to ensure the existence and uniqueness of solutions. This article deals with only the NECEEM case, but the estimation methods to be introduced also work for other KCE cases where there are two prominent peaks in the signal, such as ppKCE \cite{vasskrylov2016ccdr} or MASKE \cite{okhonin2010maske}.

The NECEEM initial conditions $\mIC(x) = c(0,x) = \bar{c}\cdot\vr(x/l)$ represent the concentration profiles of the injected plugs, where $\bar{c}= (\Leq, \Teq, \Ceq)\in\R_+^3$ denotes the initial equilibrium concentrations (note that $\Kd = \Leq\Teq/\Ceq$), furthermore $\vr:\R_+\mto\R^3$ is a vector of asymmetric Gaussian density functions, and $l$ is the theoretical ``length'' of the injected plugs. The left boundary condition vanishes $c(t,0) = 0$, while the right one is also a vanishing Neumann boundary condition $\dcdx(t,\xdet) = 0$, for computational purposes. See our previous article for further details \cite{vasskrylov2016ccdr}.

The signal is measured at the detector as the superposition of the ligand and complex concentrations
\[ \mS(t) = \mS[k](t) = \mS[k,\vg](t) \libeq (L+C)(t, \xdet) \]
where the signals may be considered to be parametrized by the rate constants $k$ and the asymmetric Gaussian plug parameters
\[ \vg = (\mu_L,\ \vs_L^1,\ \vs_L^2,\ h_L,\ \mu_T,\ \vs_T^1,\ \vs_T^2,\ h_T,\ \mu_C,\ \vs_C^1,\ \vs_C^2,\ h_C) \]
which denote the center, the left and right standard deviations, and the height of the injected plugs (dependent on the initial equilibrium concentrations).

A particular simplification of NECEEM is when $R(c) = (\koff C,\ \koff C,\ -\koff C)$ \cite{okhonin2004nonequilibrium}, for which explicit solutions have been derived when the densities are symmetric Gaussians \cite{vasskrylov2016ccdr}, but the formulas shall prove to be useful nevertheless in deriving our estimates for the parameters $k$ and $\vg$ from a given signal. Denoting
\[ F[k](t,x)\libeq \vt(t)\ \me^{-kt}\ \vr_G[v t,\ 2 D t](x) \]
\[ \vr_G(x)\libeq \vr_G[\mu, \vs^2](x)\libeq \frac{1}{\sigma\sqrt{2\pi}}\ \mathrm{exp}\left(-\frac{(x-\mu)^2}{2\sigma^2}\right),\ \ \ \vt(t)\libeq\ \begin{cases}
    1\ \ \mathrm{if}\ t\geq 0 \\
    0\ \ \mathrm{otherwise}
    \end{cases} \]
and observing that for some $\mu_0, \vs_0 >0$ and $\mu\libeq \mu_0 l,\ \vs\libeq\vs_0 l$, we have
\[ \mIC(x) = \bar{c}\ \vr_G[\mu_0, \vs_0^2](x/l) = \bar{c} l\ \vr_G[\mu,\vs^2](x) \]
then via the convolution property of Gaussians we get
\[ C(t,x) = (\mIC_C\ast F_C[\koff](t,\cdot))(x) = l \Ceq\ \vt(t)\ \me^{-\koff t}\ \vr_G[\mu + v_C t,\ \vs^2 + 2 D_C t](x). \]
The $L$ and $C$ concentrations over spacetime are the superpositions of an equilibrium and dissipation term
\[ L(t,x) = (\mIC_L\ast F_L[0](t,\cdot))(x) + \koff(C\ast F_L[0])(t,x) = l\Leq\ \vr_G[\mu+ v_L t,\ \vs^2 + 2 D_L t](x)\ + \]
\[ +\ \koff l \Ceq \int_0^t \me^{-\koff\tau} \vr_G[\mu + v_L t + (v_C-v_L)\tau,\ \vs^2 + 2 D_L t + 2(D_C-D_L)\tau](x)\ \md\tau. \]
The formula for $T$ is analogous.

\newpage
\subsection{Estimation with the Area Method} \label{s0202}

\begin{wrapfigure}{r}{0.5\textwidth}
\begin{center}
\vspace{-0.6cm}
\includegraphics[width=0.5\textwidth]{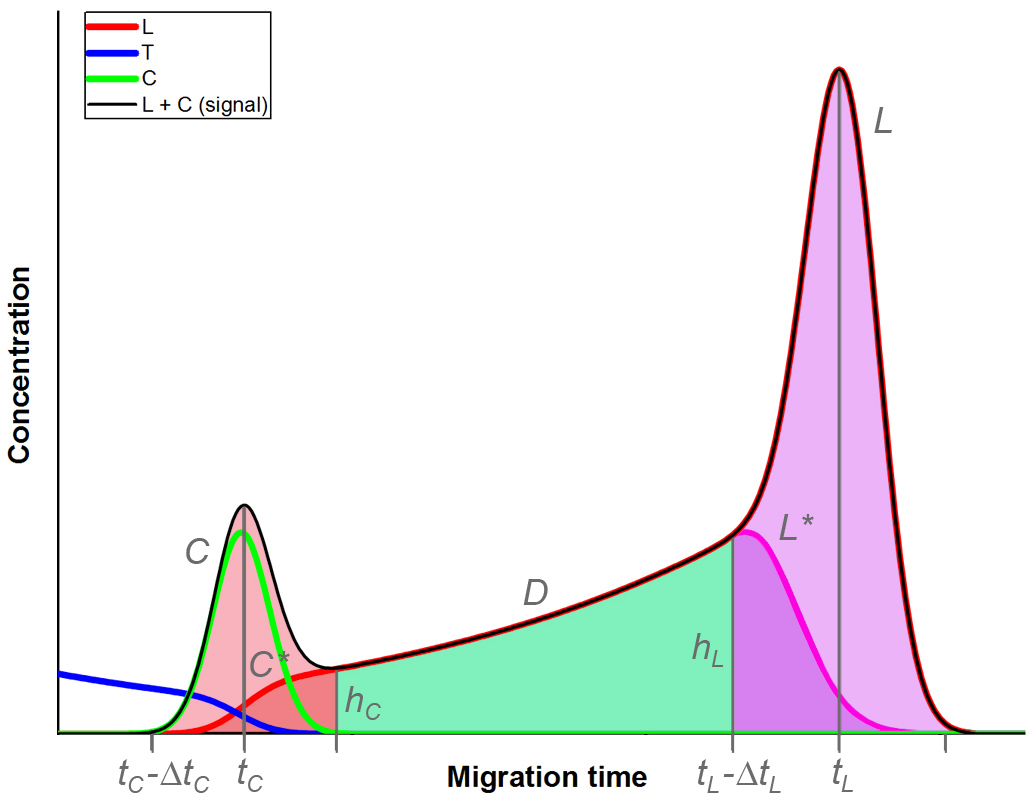}
\end{center}
\vspace{-0.15cm}
\caption{The areas defined under the signal, the three main ones being the $C$-peak, the dissociation bridge $D$, and the $L$-peak.}\label{s020201}\end{wrapfigure}
The estimation of kinetic rate constants via integration of the signal began in our lab with the work of Berezovski and Krylov \cite{berezovski2002nonequilibrium, krylov2003non}, which gave an estimate for the equilibrium binding and dissociation constants in NECEEM ($K_{\mathrm{b}}$ and $\Kd$), defined in terms of the areas over certain subintervals of the signal domain. Okhonin et al. \cite{okhonin2004nonequilibrium} built upon this work to derive an estimate for $\koff$, through the simplification of NECEEM, discussed in the previous section, but for rectangular plugs. This work was expanded to ppKCE with an estimate for $\kon$ as well \cite{okhonin2006plug}. Cherney et al. gave estimates for both $\kon$ and $\koff$ through a similar area method for NECEEM \cite{cherney2011method} and the simplified KCE case of MASKE \cite{cherney2011slow}. See also \cite{kanoatov2014extracting, kanoatov2015using, kanoatov2016systematic}.

Figure \ref{s020201} is a self-explanatory depiction of the area quantities used in the formulas for calculating $\kon$ and $\koff$. The only question that remains is how to calculate $\delt_L$ and $\delt_C$. If the signal is quite noisy, the definitions may vary, but typically they run to $1\%$ of the right end of the $L$ and the left end of the $C$ peaks, respectively. Denoting the area operator as $\mA$, the magenta and red areas -- under the ends of the dissipation tails $L^*,\ C^*$ -- have been shown to be approximately \cite{cherney2011method}
\[ \mA(L^*)\approx \frac23 h_L \delt_L,\ \mA(C^*)\approx \frac23 h_C \delt_C. \]
Defining the auxiliary constants
\[ R_1\libeq \frac{\mA(L) - \mA(L^*)}{\mA(C) + \mA(D) + \mA(L^*)},\ \ R_2\libeq \frac{\mA(C) + \mA(D) + \mA(L^*)}{\mA(C) - \mA(C^*)} \]
the dissociation constant can be approximated as
\[ \Kd \approx \frac{1000(\Tini (1 + R_1) - \Lini)}{1 + \frac{1}{R_1}} \]
where $\Lini,\ \Tini$ are the initial pre-equilibrium concentrations. The kinetic rate constants can then be approximated as
\[ \koff \approx \frac{1}{t_C}\ln(R_2),\ \ \kon = \frac{\koff}{\Kd}. \]

\subsection{The Deconvolution Operator} \label{s0203}

As we have seen earlier, the solutions of the simplified NECEEM system are expressed in terms of the function convolution operator, so in order to approximate the Gaussian parameters describing the initial concentration profiles, one may already suspect that some sort of deconvolution must be performed. For the sake of clarity and efficiency of presentation, we hereby introduce an operator for the linear least squares method performed on discrete matrices sampled from theoretically exact functions.

We consider a bivariate $F:\R^2\mto\R$ and two univariate functions $g, h:\R\mto\R$, the first of which is unknown, related by $(F(t,\cdot)\ast g)(x) = h(t,x)$. Denote the deconvolution operator as $g\libeq\Gamma(F,h)$, if such a $g$ indeed exists and it is unique. To resolve this in the discrete sense, we sample these functions on a spacetime grid defined by $t_0,\ldots,t_I$ and $x_0,\ldots,x_J$, in the intervals $I_t\libeq [t_0, t_I],\ I_x\libeq [x_0, x_J]$. Specifically, we are interested in deconvolution at $x_J$ (which typically corresponds to the detector location $\xdet$). Thus at some $t_i$, we have
\[ h(t_i, x_J) = ((F(t_i),\cdot)\ast g)(x_J) = \int_0^{x_J} F(t_i, x_J - x) g(x)\ \md x \approx \sum_{j=1}^J F(t_i, x_J - x_j) g(x_j) \delx \]
where $\delx = x_J/J$, implying a potentially over/under-determined system of linear equations, which can be written in matrix form as
\[
\begin{bmatrix}
    F(t_0, x_J - x_1) & \dots & F(t_0, x_J - x_J) \\
    \vdots & \ddots & \vdots \\
    F(t_I, x_J - x_1) & \dots & F(t_I, x_J - x_J)
\end{bmatrix}
\cdot
\begin{bmatrix}
    g(x_1) \\
    \vdots \\
    g(x_J)
\end{bmatrix}
=
\frac{J}{x_J}
\cdot
\begin{bmatrix}
    h(t_0, x_J) \\
    \vdots \\
    h(t_I, x_J)
\end{bmatrix}.
\]
Clearly, exact equality may not be attainable, meaning no such vector of values $(g(x_1),\ldots, g(x_J))$ is likely to exist, though a least squares approximation does. Denoting the matrices as $\widehat{F},\ \widehat{g},\ \widehat{h}$, the vector $v\in\R^J$ that minimizes the error $\|\widehat{F}v - \widehat{h}\|_2$ in the Euclidean norm, is given by
\[ \widehat{\Gamma}(F,h) = \widehat{\Gamma}_{I_t, I_x}(F,h) = \widehat{\Gamma}_{I, J}(F,h)\libeq (\widehat{F}^{T}\!\widehat{F})^{-1} \widehat{F}^{T}\widehat{h} \approx \widehat{g}. \]
Clearly, this explicitly-defined discrete deconvolution operator approaches the continuous one in the limit (in some appropriate operator metric), as the spacetime grid becomes denser.

\newpage
\section{Estimation Methods} \label{s03}

\subsection{Estimation of $\koff$} \label{s0301}

As described earlier in Section \ref{s0201}, the signal $\mS[k,\vg]$ in the simplified NECEEM case is given by the formula
\[ \mS[k,\vg](t) = L(t,\xdet) + C(t,\xdet) = L_{\meq}(t,\xdet) + L_{\mdis}(t,\xdet) + C(t,\xdet) = \]
\[ = l\Leq\ \vr_G[\mu+ v_L t,\ \vs^2 + 2 D_L t](x)\ + \]
\[ +\ \koff l \Ceq \int_0^t \me^{-\koff\tau} \vr_G[\mu + v_L t + (v_C-v_L)\tau,\ \vs^2 + 2 D_L t + 2(D_C-D_L)\tau](x)\ \md\tau\ + \]
\[ +\ l \Ceq\ \vt(t)\ \me^{-\koff t}\ \vr_G[\mu + v_C t,\ \vs^2 + 2 D_C t](x). \]

The third term in the sum is $C(t,\xdet)$, which is dominated by the Gaussian factor only near the $C$-peak, while along the dissociation bridge the exponential factor $\me^{-\koff t}$ dominates. Therefore heuristically $C(t,\xdet)\approx \lambda \me^{-\koff t}$, with some $\lambda>0$ along the dissociation bridge. The second term $L_{\mdis}(t,\xdet)$ can actually be expressed similarly, since
\[ \ddt L_{\mdis}(t,\xdet) = \koff C(t,x)\approx \koff\lambda \me^{-\koff t} \]
implying on the dissociation bridge that $L_{\mdis}(t,\xdet)\approx -\lambda \me^{-\koff t}$. Lastly, the first term in the above superposition is the $L$-peak, which is roughly constant along the bridge. Thus we can conclude that the signal along the bridge decays approximately exponentially, with a rate of $\koff$.

This implies the following method for approximating $\koff$. Take a subinterval of the bridge which does follow an exponential (such as by taking the natural logarithm of the signal, and finding an interval where its second derivative is near-zero). Then take the logarithm of the signal on that subinterval, and perform a linear least squares approximation. The resulting slope is our approximation of $\koff$.

\newpage
\subsection{Estimation of $\vg$ and $\kon$} \label{s0302}

Assuming that the initial concentration profiles of the injected plugs are asymmetric Gaussian density functions, with the vector of parameters
\[ \vg = (\mu_L,\ \vs_L^1,\ \vs_L^2,\ h_L,\ \mu_T,\ \vs_T^1,\ \vs_T^2,\ h_T,\ \mu_C,\ \vs_C^1,\ \vs_C^2,\ h_C)\]
from Section \ref{s0201}, the three profiles are entirely determined by these twelve parameters. In order to approximate them, however, a deconvolution must be performed on ideal portions of the signal.

Our first aim is to identify subintervals of the signal support, where the deconvolutions may be ideally performed. One may observe heuristically, that the bottom one third of the left-half of the $C$-peak values arise solely from the values of $C$, unaffected by $L$. Denote this interval $I_C$. Similarly, the bottom one third of the right half of the $L$-peak values arise solely from the equilibrium values of $L$, unaffected by $C$ or the dissipation values of $L$. Denote this interval $I_L$. Also denote $I_x\libeq [0,\xdet]$. The following hold over these intervals
\[ L(t,\xdet)\approx (F_L[0](t,\cdot)\ast \mIC_L)(\xdet),\ \ C(t,\xdet) = (F_C[\koff](t,\cdot)\ast \mIC_C)(\xdet) \]
defined earlier as
\[ F_L[0](t,x) = \vr_G[v_L t,\ 2 D_L t](x),\ \ F_C[\koff](t,x) = \me^{-\koff t} \vr_G[v_C t,\ 2 D_C t](x). \]

In order to perform the deconvolutions according to Section \ref{s0203}, we first discretize the spacetime intervals $I_L\times I_x,\ I_C\times I_x$. Then approximations to the initial concentration profiles are given by the discrete deconvolutions
\[ \widehat{\mIC}_L = \widehat{\Gamma}_{I_L, I_x}(F_L[0],\ L(\cdot,\xdet)),\ \ \ \widehat{\mIC}_C = \widehat{\Gamma}_{I_C, I_x}(F_L[\koff],\ C(\cdot,\xdet)). \]

The corresponding parameters in $\vg$ can be deduced from these two profiles, as follows. The centers and standard deviations, are the peak locations and the inflection points of the asymmetric Gaussian initial condition profiles $\widehat{\mIC}_L$ and $\widehat{\mIC}_C$. Furthermore, the initial equilibrium concentrations can be calculated from the initial condition heights as follows
\[ \Leq = h_L \sqrt{2\pi}\ \frac{\vs_L^1 + \vs_L^2}{2},\ \ \Ceq = h_C \sqrt{2\pi}\ \frac{\vs_C^1 + \vs_C^2}{2},\ \ \Teq = 1000^2 \Tini l - \Ceq. \]
The center and standard deviations of the $T$ initial condition (injected plug concentration profile) are approximated as
\[ c_T \approx \frac{c_L + c_C}{2},\ \ \vs_T^1 \approx \frac{\vs_L^1 + \vs_C^1}{2},\ \ \vs_T^2 \approx \frac{\vs_L^2 + \vs_C^2}{2}. \]

Lastly, according to Section \ref{s0201}, the kinetic rate constant of complex formation can be estimated via the exact relationship $\kon = \koff\Ceq/(\Leq\Teq)$. Thus the error in estimating the concentration components of $\bar{c}$ accumulates in this $\kon$ estimate, as analyzed in the next section.

\newpage
\section{Computational Results} \label{s04}

\subsection{Error Comparison between the Area and Our Method} \label{s0402}

\begin{figure}[H]
\centering
\includegraphics[width=375pt]{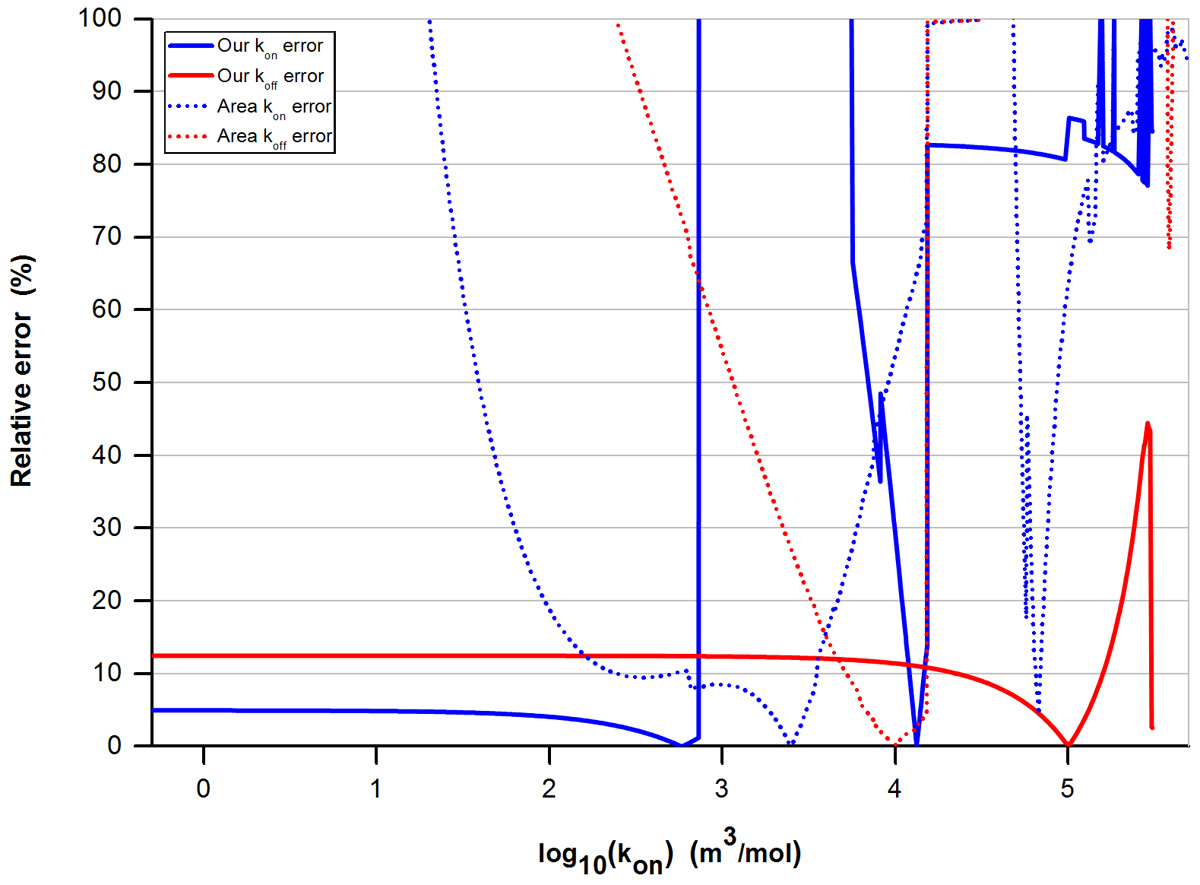}
\caption{Kinetic parameter estimation errors with the area and our method.}
\label{s040201}
\end{figure}

We performed a computational comparison on Figure \ref{s040201} between the area method (Section \ref{s0202}) and our method (Section \ref{s03}), over a set of kinetic rate constants with constant $\Kd$ and varying $\kon$, in a neighborhood of three orders of magnitude, with $1000$ uniformly distributed random sample points, centered at
\[ \kon = 500\mmcmol,\ \koff = 0.001,\ v = (0.22, 3.33, 0.48)\times 10^{-3}\mmsVsec,\ D = (7, 7, 7)\times 10^{-11}\mmssec \]
\[ \bar{c} = (2.1, 18.1, 1.9)\times 10^{-7}\mmolmc,\ l = 0.01\mmet,\ t_{\max} = 1200\msec,\ \xdet = 0.2\mmet. \]
Figure \ref{s040201} is plotted on a logarithmic scale in the independent variable $\kon$, with the center of the sample interval being $\log_{10}(500) = 2.6990$. $\Kd$ is kept constant at a value of $2\times 10^{-6}\mmolmc$. The relative error is calculated as the ratio of the absolute distance between the original and the estimated value, and the original value. See our software \cite{so002}.

Upon plotting the relative errors, it is perhaps most important to observe that all four estimation methods are near-exact for particular $\kon$ values -- meaning, the relative error nearly vanishes -- which confirms the reliability of the methods in this sense. The first local minimum in the log-scale sample, for our $\kon$ estimation method, occurs at $2.7656$ with a relative error of $0.03\%$, for our $\koff$ method at $5.0034$ with $0.20\%$, for the area $\kon$ method at $3.3975$ with $0.08\%$, and for the area $\koff$ method at $4.0210$ with $0.23\%$. Interestingly, beyond these critical values, the errors for all four methods become unpredictable, making them all unreliable for higher $\kon$ values.

Thus based on Figure \ref{s040201}, we conjecture that both our $\kon$ and $\koff$ estimates are reliable up to certain $\kon$ values, the estimation of which we leave as an open problem.

\subsection{Error Analysis of the Initial Concentration Profiles} \label{s0403}

To get an idea of the relative errors incurred by the estimation of the $\vg$ parameter vector, characterizing the asymmetric Gaussian plug concentration profiles, we perform the estimation of this vector using exact (non-estimated) $\kon$ and $\koff$ values, on the same logarithmic interval as before, while keeping the original $\vg$ coordinates constant at the above values.

\begin{figure}[H]
\centering
\includegraphics[width=375pt]{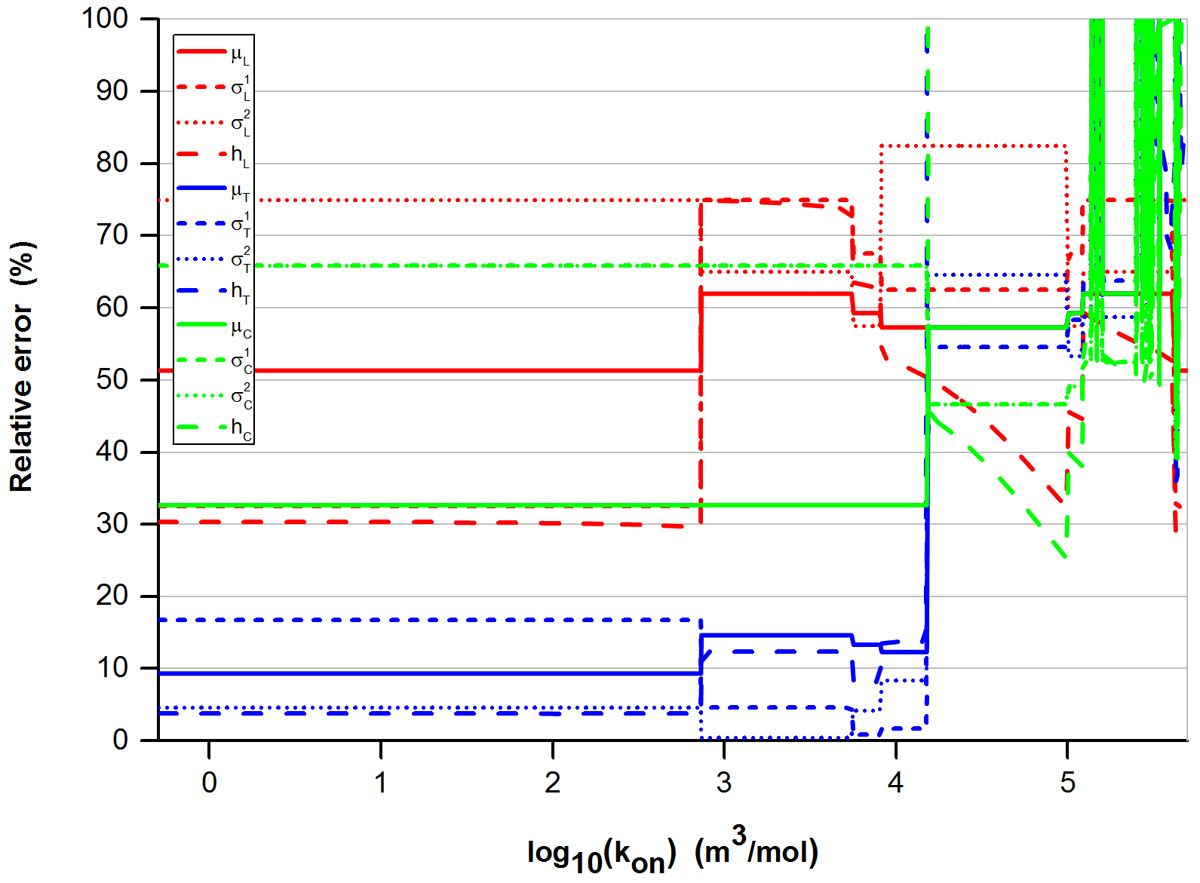}
\caption{Plug concentration parameter estimation errors.}
\label{s040301}
\end{figure}

The resulting Figure \ref{s040301} shows the remarkable constancy of the relative errors between the original and the estimated $\vg$ values up to certain critical values (interestingly, $h_L$ is not quite constant). These critical values occur in our sample for the $L$- and $T$-plug parameters $\mu_L, \vs_L^1, \vs_L^2, h_L, \mu_T, \vs_T^1, \vs_T^2, h_T$ at the $\log_{10}(\kon)$ value of $2.8618$, while the $C$-plug errors for $\mu_C$ and $\vs_C^2$ remain constant a bit longer, and those for $\vs_C^1$ and $h_C$ are off the chart or erratic. This behavior is partly due to the nature of the estimation method presented in Section \ref{s0302}, and partly to our MATLAB implementation \cite{so002}. Note that the error that accumulates in $\vg$ towards estimating $\kon$ remains reasonable, according to the previous section.

\newpage

\newpage
\section{Concluding Remarks} \label{s05}

We have introduced some estimation methods, based on linear regression, for kinetic constants of macromolecules, as well as for the parameters of the initial concentration profiles of the injected plugs, within the experimental framework of Kinetic Capillary Electrophoresis. As demonstrated through computational testing, our rate constant estimation proved to be more reliable, up to some conjectured upper bound, than our former integration-based method. Our initial concentration profile parameter estimation method may curiously likewise be conjectured to be reliable up to some upper bound.

The proof of these conjectures could be a matter of future effort, however, it may not be particularly worthwhile, considering that the utility of the rate constant and concentration parameter estimates is merely in their role as an initial starting point for our computational resolution of the KCE inverse problem \cite{so002}, to be detailed in our upcoming papers. While a reliable starting point is preferable for the inversion -- i.e. one ``typically close enough'' to the sought solution -- it is not imperative. Nevertheless, an estimation method that is known to be ``robust'' according to sufficient testing, can serve as a check on whether the inversion diverges -- i.e. the optimization algorithm performing the minimization of the error function, defined between the target and the simulated signals.

\newpage
\bibliographystyle{abbrv}
\bibliography{mybib2}
\addcontentsline{toc}{section}{\textbf{References}}

\end{document}